\documentstyle[preprint,aps]{revtex}
\begin{document}

\draft
\title{A model of marginal Fermi liquid superconductor at 
two dimensions }
\author{Yu-Liang Liu and T. K. Ng}
\address{Department of Physics, Hong Kong University of Science and 
Technology, Clear Water Bay, Kowloon, Hong Kong, People's Republic of China}
\maketitle
\begin{abstract}

    In this paper we study a model of s-wave marginal Fermi liquid 
superconductor at two dimensions.  Besides the usual 
Bardeen-Cooper-Schrieffer(BCS) point attractive interaction, the fermions in 
our system also interact {\em via} a long range repulsive interaction 
$v(q)=g_{D}/q^2$. We find that although superconductivity is not destroyed 
by the long-ranged repulsion interaction (demonstrated by the existence of 
Meissner effect), nevertheless the BCS pairing function $\Delta_{BCS}$ is 
strongly suppressed resulting in power-law correlation $<\Delta_{BCS}(\vec{x})\Delta_{BCS}(0)>
\sim|\vec{x}|^{-2\gamma}$. This observation is explained within a general picture of marginal Fermi liquid.

\end{abstract}
\vspace{1cm}


\newpage

Since the discovery of high-$T_c$ superconductors, there has been enormous 
interests in the study of non-Fermi liquid theories in two 
dimensions\cite{1,2,3}. The normal state of high-$T_c$ superconductors 
shows rich physical behaviors which seems to deviate from usual Landau 
Fermi liquid theory predictions. For example, in the optimally doped regime, 
resistivity shows a linear temperature dependence, whereas in underdoped 
regime, the resistivity shows a power-law temperature dependence, and with 
a pseudo-gap in low energy excitation spectrum. 
Although there exists a lot of theoretical studies on microscopic theories of 
non-Fermi liquid normal states, at present there exists very few systematic 
investigation into how the superconducting state is affected by non-Fermi 
liquid behaviours. 

   In this paper we study a microscopic model of s-wave marginal Fermi 
liquid superconductor at two dimensions. The Hamiltonian of our model 
consists of two parts, $H=H_{mfl}+H_{BCS}$, where
\begin{mathletters}
\label{ham}
\begin{equation}
\label{hmf}
H_{mfl}=\displaystyle{ \sum_{\vec{k},\sigma}\epsilon(k)\psi^{\dagger}_{\sigma}(\vec{k})
\psi_{\sigma}(\vec{k})+
\frac{1}{2L^2}\sum_{|\vec{q}|\leq q_{c}}v(q)\rho(\vec{q})\rho(-\vec{q})},
\end{equation}
and
\begin{equation}
\label{hbcs}
H_{BCS}=-\displaystyle{ \int d^{2}x[\Delta(\vec{x},t)\psi^{\dagger}_{\downarrow}(\vec{x},t)
\psi^{\dagger}_{\uparrow}(\vec{x},t)+
\Delta^{*}(\vec{x},t)\psi_{\uparrow}(\vec{x},t)
\psi_{\downarrow}(\vec{x},t)-{1\over{U}}|\Delta(\vec{x},t)|^2]}
\label{1}\end{equation}
\end{mathletters}
where $\epsilon(k)=\frac{k^{2}}{2m}-\mu$, $\rho(\vec{x})=\sum_{\sigma}
\psi^{\dagger}_{\sigma}(\vec{x})\psi_{\sigma}(\vec{x})$ is the electron
density operator, $\psi_{\sigma}(\vec{x})(\psi^{\dagger}_{\sigma}(\vec{x}))$ 
are electronic annihilation (creation) operators with spin $\sigma=\uparrow, 
\; \downarrow$, $L^2$ is the volume of the system,
$q_{c}<<k_F$ is a high momentum cutoff, $v(q)=g_{D}/q^2$, and 
$\Delta(\vec{x},t)(\Delta^{*}(\vec{x},t))$ are Hubbard-Stratonovich fields 
introduced to decouple the (attractive) electron-electron interaction $U$. 
$U$ is a coupling constant of an unknown mechanism which leads to electron 
pairing. Note that in usual BCS mean-field theory, $\Delta(\vec{x},t)\rightarrow
\Delta_{BCS} = -U<\psi_{\uparrow}(\vec{x})\psi_{\downarrow}(\vec{x})>$, where $\Delta_{BCS}$ is the 
mean-field BCS gap. The repulsive interaction potential $v(q)$ among the 
electrons may originate from an U(1) gauge field such as in a t-J 
model\cite{3,8}, or a vortex-vortex interaction where each vortex is 
combined to one electron\cite{9}.
 
The Hamiltonian $H_{mfl}$ has been carefully studied\cite{5} where it is 
known that it describes a state with Fermi-liquid like particle-hole and 
collective excitation spectrums, but with nevertheless a vanishing Fermi 
surface discontinuity $z\rightarrow0$\cite{5} because of orthogonality 
catastrophe effect associated with induced long-ranged density fluctuations 
when an additional electron is introduced\cite{5} (marginal Fermi liquid 
behaviour). Notice that in the absence of interaction term $v(q)$, 
the system would be a regular s-wave superconductor at two dimensions, 
described by a constant pairing function $\Delta_{BCS}$ in mean-field theory. 
Here we shall examine whether superconductivity is destroyed by the marginal 
Fermi liquid behaviour, and if it survives, whether and how its behaviour 
will be modified. Note that even though it is believed that in the high-$T_c$ 
cuprates the electron pairing function has a d-wave symmetry, 
interesting conclusions may still be drawn from studying this simple 
model as we shall see in the following. 

To solve this model Hamiltonian we first introduce another 
Hubbard-Stratonovich field $\phi(q,t)$ to decouple the repulsive 
electron-electron interaction. In terms of $\phi$ and $\Delta$ fields, 
the generating functional of the system can be written as,
\begin{mathletters}
\label{action}
\begin{equation}
\label{a1}
Z=\displaystyle{ \int D\Psi D\Psi^{\dagger}
D\phi e^{\displaystyle{i \int^{T}_{-T} dt[\frac{1}{L^{2}}\sum_{|\vec{q}|\leq q_c}
\frac{\phi(\vec{q},t)\phi(-\vec{q},t)}{2v(q)}+{\cal L}(t)]}}},
\end{equation}
where $T\rightarrow\infty$,
\begin{equation}
\label{a2}
{\cal L}(t)=\displaystyle{ \int d^{2}x\Psi^{\dagger}(\vec{x},t)
\hat{G}^{-1}(\vec{x},t,[\phi])\Psi(\vec{x},t)},
\end{equation}
and
\begin{equation}
\label{a3}
\hat{G}^{-1}(\vec{x},t,[\phi])=\displaystyle{ 
i\partial_{t}+(\frac{\nabla^{2}}
{2m}+\mu)\sigma^{z}-\tilde{\Delta}(\vec{x},t)-\tilde{\phi}(\vec{x},t)},
\end{equation}
\end{mathletters}
where $\sigma^{z}$ is a Pauli matrix, 
$\Psi(\vec{x},t)=\left(\begin{array}{c} \psi_{\uparrow}(\vec{x},t)\\
\psi^{\dagger}_{\downarrow}(\vec{x},t)\end{array}\right)$, 
$\tilde{\Delta}(\vec{x},t)=\left(
\begin{array}{cc} 0, & \Delta(\vec{x},t)\\ 
\Delta^{*}(\vec{x},t), & 0 \end{array}\right)$,
and $\tilde{\phi}(\vec{x},t)=\left(\begin{array}{cc} \phi(\vec{x},t), & 0\\
0, & -\phi(\vec{x},t)\end{array}\right)$. Note that the Lagrangian is now 
quadratic in electron fields after the HS bosonic fields $\Delta(\vec{x},t)$ 
and $\phi(\vec{x},t)$ are introduced. 

   We shall solve the problem in two steps: (1)for each configuration 
of $\phi(\vec{x},t)$ field, we first solve the self-consistent BCS mean-field 
equations
\begin{eqnarray}
\label{mfbcs}
 && \hat{G}^{-1}(\vec{x},t,[\phi]) G(\vec{x},\vec{x}',t,t',[\phi]) = \delta(\vec{x}-\vec{x}')
\delta(t-t'),   \\   \nonumber
 && \Delta(\vec{x},t;[\phi]) = -U<\psi_{\uparrow}(\vec{x},t;[\phi])
\psi_{\downarrow}(\vec{x},t;[\phi])>_0,
\end{eqnarray}
approximately using an Eikonal expansion\cite{11,11'} or 
functional bosonization\cite{10}
approach, where $<A>_0$ is the expectation value of operator $A$ at fixed 
configuration $\phi(\vec{x},t)$, and (2)with the approximate Green's function 
$G(\vec{x},\vec{x}',t,t';[\phi])$ obtained for each configuration $\phi(\vec{x},t)$, 
we can construct an approximate action $S_{eff}[\phi]$ in terms of 
$\phi(\vec{x},t)$ field only in the Eikonal expansion. The ground state 
expression value
$<G(\vec{x},t)>$ and other higher-order correlation 
functions can be evaluated as
\begin{equation}
\label{phy}
<G(\vec{x}-\vec{x}',t-t')>=<G(\vec{x},\vec{x}',t,t';[\phi]>_{\phi}=
{\int{d}\phi{G}(\vec{x},\vec{x}',t,t';[\phi])
e^{-S_{eff}[\phi]}\over\int{d}\phi{e}^{-S_{eff}[\phi]}},
\end{equation}
etc. Notice that Eikonal expansion can be used as a systematic expansion technique 
in powers of $q_c/k_F$\cite{11,11'} for fermions interacting with scalar 
potential $v(q)$. In particular, the marginal Fermi liquid behaviour 
can be recovered correctly to leading order in Eikonal 
expansion\cite{11,11',6}. The new ingredient we introduce here is to 
include the BCS mean-field theory in a self-consistent way into the 
formalism. In the following we shall give some brief mathematical outlines 
on how the Eikonal expansion is carried out. First we review the case when 
the BCS term is absent (normal marginal Fermi liquid). 

   In this case the matrix Green's function 
$G(\vec{x},\vec{x}',t,t';[\phi])$ becomes a 
number function. In the Eikonal expansion, we write $G$ as\cite{10,6}
\begin{equation}
G(\vec{x},\vec{x}',t,t';[\phi])={1\over TL^{2}}\sum_{\vec{k},\omega}
g_0(\vec{k},\omega)e^{-i[\omega{t}-\vec{k}.\vec{x}+Q_{\vec{k}}
(\vec{x},t;[\phi])]}e^{i[\omega{t}'-\vec{k}.\vec{x}'+Q_{\vec{k}}
(\vec{x}',t';[\phi])]},
\end{equation}
where $g_0(\vec{k},\omega)=(\omega-\epsilon(k))^{-1}$, $\epsilon(k)=k^{2}/
(2m)-\mu$,
and the function $Q_{\vec{k}}
(\vec{x},t;[\phi])$ is obtained approximately in an Eikonal expansion. 
To leading order in $q_c/k_F$ we obtain
\[
Q_{\vec{k}}(\vec{x},t;[\phi])=-{1\over TL^2}\sum_{|\vec{q}|<q_c,\omega}
e^{i(\omega{t}-\vec{q}.\vec{x})}{\phi(\vec{q},\omega)\over\omega-
{\vec{k}.\vec{q}\over{m}}}.  
\]
For the special case of a $1/q^2$ potential in two dimension, it turns 
out that the infra-red singularity which leads to marginal-Fermi liquid 
behaviour comes from plasma oscillations with energy 
$\omega\sim\Omega_0=\sqrt{g_{D}\rho_{0}/m}$, 
where $\rho_{0}$ is the electron density\cite{5,6}. 
As a result one may neglect the 
$\vec{q}.\vec{k}/m$ term in $Q_{\vec{k}}$ as far as the leading 
infra-red singularity is concerned, and we obtain
\begin{equation}
\label{gm}
G(\vec{x},\vec{x}',t,t';[\phi])=G_0(\vec{x}-\vec{x}',t-t')
e^{i(Q(\vec{x},t)-Q(\vec{x}',t'))}, 
\end{equation}
where $Q(\vec{x},t)\sim-{1\over{T}L^2}\sum_{\vec{q},\omega}
e^{i(\omega{t}-\vec{q}.\vec{x})}
{\phi(\vec{q},\omega)\over\omega}$, 
$G_0(\vec{x},t)$ is the usual free fermion 
Green's function. An Eikonal expansion can also 
be performed for the effective action $S_{eff}(\phi)$. 
To leading order in $q_c/k_F$, $S_{eff}[\phi]$ is just 
the RPA effective action
\[
S_{eff}[\phi]={1\over2TL^2}\sum_{q<q_c,\omega}
|\phi(\vec{q},\omega)|^2[v(q)^{-1}-\chi_0(q,\omega)],  
\]
where $\chi_0(q,\omega)$ is the usual Linhard function. This result can be 
understood by noting that (i)higher order terms in power of $\phi$ 
involves sums over extra momentum $q<q_c$, and generate higher powers in 
$q_c/k_F$, and (ii)the singular behaviour at small $q$ which is present in 
quadratic term will be absent in higher order terms. Notice also that the 
marginal Fermi liquid behaviour originates from electron-plasmon coupling 
in the present model\cite{5}. Therefore, as far as the marginal Fermi liquid 
behaviour is concerned, the effective action may be further simplified by 
keeping only the plasma-pole contributions\cite{5} to $S_{eff}[\phi]$. This 
is in contrast to many other models of marginal Fermi liquids where the 
marginal Fermi liquid behaviour is generated from coupling of electrons to 
anomalous low energy particle-hole excitations\cite{2,3}.

   With Eq.\ (\ref{gm}), it is easy to show that the corresponding 
matrix Green's function in the presence of BCS attractive interaction 
which keeps the leading infra-red singularity has form, 
\begin{equation}
\label{gbcs}
G(\vec{x},\vec{x}',t,t';[\phi])=\left(\begin{array}{cc} 
G_0^{BCS}e^{i(Q(\vec{x},t)-Q(\vec{x}',t'))}, 
& F_0^{BCS}e^{i(Q(\vec{x},t)+Q(\vec{x}',t'))} \\
F_0^{BCS*}e^{-i(Q(\vec{x},t)+Q(\vec{x}',t'))}, & 
G_0^{BCS*}e^{-i(Q(\vec{x},t)-Q(\vec{x}',t'))}
\end{array}\right).
\end{equation}
and with corresponding pairing function
\begin{equation}
\label{pair}
\Delta(\vec{x},t;[\phi])={U\Delta\over2L^2}
\sum_{\vec{k}}{1\over{E}(k)}e^{-2iQ(\vec{x},t)}
\end{equation}
where $G_0^{BCS}\equiv G_{0}^{BCS}(\vec{x}-\vec{x}',t-t')$ and 
$F_0^{BCS}\equiv F_{0}^{BCS}(\vec{x}-\vec{x}',t-t')$ 
are the usual BCS normal and anomalous 
Green's functions in the absence of $\phi$ field, respectively, with usual 
Fourier transform,
\begin{eqnarray}
G_0^{BCS}(\vec{k},i\omega) & = & 
{i\omega+\epsilon(k)\over(i\omega)^2-E(k)^2},  \\  \nonumber
F_0^{BCS}(\vec{k},i\omega) & = & {-\Delta\over(i\omega)^2-E(k)^2},
\end{eqnarray}
where $E(k)=\Delta^2+\epsilon(k)^2$, $\Delta$ is determined by the usual 
BCS mean-field equation $1= \frac{U}{2L^2}\sum_{k}\frac{1}{E_{k}}$ 
(see (\ref{pair})). We shall assume $\Delta<<\Omega_0$ in the following
discussions.
 
  As in the normal state, to leading order in $q_c/k_F$, the effective 
action $S_{eff}[\phi]$ of the system in the presence of BCS attractive 
interaction is the quadratic action of superconductors for scalar 
potentials, 
\begin{eqnarray}
S_{eff}[\phi] &=& \displaystyle{ 
\frac{1}{TL^{d}}\sum_{q\leq |q_{c}|,\Omega}\left[
\frac{1}{2v(q)}-\chi_{0}(q,\Omega)+\Gamma(q,\Omega)\right]|\phi(q,\Omega)|^{2}}
\nonumber \\
\Gamma(q,\Omega) &=& \displaystyle{ \frac{1}{L^2}\sum_{k}
\frac{4\Delta u_{k}v_{k}\xi^{2}_{q}\theta(-\epsilon(k))}
{(\Omega^{2}-
\varepsilon^{2}_{k}(q))^{2}+4\Delta^{2}\xi^{2}_{q}}} \label{12} \\
\chi_{0}(q,\Omega) &=& \displaystyle{ \frac{1}{L^2}\sum_{k}
\frac{\xi_{q}\theta(-\epsilon(k))}{(\Omega^{2}-\varepsilon^{2}_{k}(q))^{2}+
4\Delta^{2}\xi^{2}_{q}}[(\Omega+\varepsilon_{k}(q))^{2}-2u^{2}_{k}
(\Omega^{2}+\varepsilon^{2}_{k}(q))]}.
\nonumber\end{eqnarray}
where $\varepsilon_{k}(q)=\vec{k}\cdot\vec{q}/m$, $\xi_{q}=q^{2}/(2m)$, 
$u^{2}_{k}=(1+\epsilon(k)/E(k))/2$, $v^{2}_{k}=(1-\epsilon(k)/E(k))/2$, and
$\theta(x)$ is the step function.
In the long wave-length limit $q\rightarrow 0$, $\chi_{0}(q,\Omega)$ and
$\Gamma(q,\Omega)$ are reduced to $\chi_{0}(q\rightarrow 0, \Omega)\simeq
\frac{\rho(\Delta)}{2m}(\frac{q}{\Omega})^{2}$, and $\Gamma(q\rightarrow 0,
\Omega)\simeq \frac{\Delta^{2}}{m^{2}U}(\frac{q}{\Omega})^{4}\sim 0$, 
where $\rho(\Delta)=
\frac{2}{L^2}\sum_{k}v^{2}_{k}=\rho_0$. For a $v(q)=g_{D}/q^{2}$ potential, 
the effective action becomes in the limit $q\rightarrow 0$ and $\Omega>>
k_Fq/m$,
\begin{equation}
S_{eff.}[\phi]=\frac{1}{TL^{2}}\sum_{q,\Omega}\frac{1}{2v(q)}\left(
1-\frac{\Omega^{2}_{0}}{\Omega^{2}}\right)|\phi(q,\Omega)|^{2}
\label{13}\end{equation}
where $\Omega_{0}=\sqrt{g_{D}\rho_0/m}$ is the plasma frequency.
Notice that as far as plasma oscillation 
is concerned, the effective action $S_{eff}[\phi]$ is unaffected by presence 
of supeconductivity. This is a direct manifestation of Higg's mechanism.

Using the effective action (\ref{13}), we can calculate the 
thermodynamical average
and correlation function of the pairing function and the 
normal and anomalous Green's functions by using
their expressions in (\ref{gbcs}) and (\ref{pair}), respectively. We obtain
\begin{eqnarray}
<Q(\vec{x},t)Q(\vec{x}',t')>_{\phi} &\simeq& \left\{
\begin{array}{ll} \displaystyle{
-\frac{\gamma}{4}\ln |\vec{x}-\vec{x}'|}, & |t-t'|\ll\frac{1}{
\Omega_{0}},\\
\displaystyle{
\frac{\gamma}{4}\ln(\frac{1}{q_{c}L})}, & |t-t'|\gg\frac{1}{
\Omega_{0}}, \end{array}\right. 
\end{eqnarray}
where $\gamma={g_D\over2\pi\Omega_0}$, and correspondingly,
\begin{eqnarray}
\Delta_{BCS}=<\Delta(\vec{x},t,[\phi])>_{\phi} &\simeq& \displaystyle{ 
\Delta\left(\frac{1}{q_{c}L}
\right)^{\gamma}}, \label{14} \\
<\Delta(\vec{x},t,[\phi])\Delta^*(\vec{x}',t,[\phi])>_{\phi} 
&\simeq& \displaystyle{
\Delta^{2}\left(\frac{1}{|\vec{x}-\vec{x}'|}\right)^{2\gamma}}, \nonumber \\
<\Delta(\vec{x},t,[\phi])\Delta^*(\vec{x}',t',[\phi])>_{\phi} &\simeq& 
\displaystyle{
\Delta^{2}\left(\frac{1}{q_{c}L}\right)^{2\gamma}},
\;\; |t-t'|\rightarrow\infty.
\nonumber\end{eqnarray}
 Notice that as the size of the system $L^2\rightarrow\infty$, the 
pairing function $\Delta_{BCS}$ goes to zero because of the long range 
interaction potential $v(q)$, and its (equal-time) correlation function 
shows a power-law behavior! The normal electron Green's function can also 
be calculated similarly. We obtain 
\begin{eqnarray}
\label{g1}
G(\vec{x}-\vec{x}',t-t') &\simeq& G_0^{BCS}(\vec{x}-\vec{x}',t-t')\left\{
\begin{array}{ll} \displaystyle{ \left(
\frac{1}{|\vec{x}-\vec{x}'|}\right)^{\gamma}}, 
& t\sim{t}', \\
\displaystyle{ 
\left({1\over{q}_cL}\right)^{\gamma}}, & |t-t'|>>{1\over\Omega_0}, 
\end{array}\right. 
\end{eqnarray}
   Notice that a similar expression of the one-electron Green's function is 
also obtained in the normal state with $G_0^{BCS}$ replaced by the usual 
free fermion Green's function. 

  There are two effects to be noticed. First of all, we find that the BCS 
pairing function is strongly suppressed by the long-range interaction 
potential, which seems to indicate that superconductivity is suppressed. 
Secondly, the one-electron Green's function is strongly suppressed by a 
term $(q_cL)^{-\gamma}$ at frequency $\omega<\Omega_0$, resulting in 
a 'pseudo-gap' like structure in the tunneling density of state for 
$\omega<\Omega_0$, both in the normal state\cite{5} and superconductivity 
state. Physically, this 'pseudo-gap' like structure indicates that in the 
presence of long-range repulsive interaction, it is essentially impossible 
to add (or subtract) an electron without creating density fluctuations in 
the system\cite{5}. Notice, however that although the differences between
superconducting and normal states in the normal and anomalous one-electron 
Green's functions are strongly suppressed by $v(q)$ and are hardly visible
when $\Delta<<\Omega_0$, nevertheless an order parameter $\Delta$ determined 
by the BCS mean-field equation still exists. We shall look into the meaning
of nonzero $\Delta$ more carefully by examining directly the Meissner effect in
the following.  

  Following standard linear-response theory analysis, we find in the Coulomb
gauge  
\begin{equation}
\label{lr}
\vec{J}(\vec{x},t)=\int{d}^2x'P(\vec{x}-\vec{x}',t-t')\vec{A}(\vec{x}',t')-{\rho_0\over{m}}\vec{A}(\vec{x}.t)
\end{equation}
where $\vec{A}(\vec{x},t)$ is the external (transverse) gauge field and $\vec{J}(\vec{x},t)$ is
the corresponding induced current. The response function $P(\vec{x},t)$ is given by
\begin{eqnarray}
\label{tr}
P(\vec{x}-\vec{x}',t-t') & = & <P(\vec{x},\vec{x}',t,t';[\phi])>_{\phi},  \\ \nonumber
P(\vec{x},\vec{x}',t,t';[\phi]) & = & {-i\over{m}^2}Tr\left[(\nabla_x.\nabla_{x'})G(\vec{x},\vec{x}',
t,t';[\phi])G(\vec{x}',\vec{x},t',t;[\phi])\right],
\end{eqnarray}
where $(\nabla_x.\nabla_{x'})G_1G_2=(\nabla_xG_1).(\nabla_{x'}G_2)+(\nabla_{x'}G_1).(\nabla_{x}G_2)-
(\nabla_x.\nabla_{x'}G_1)G_2-G_1(\nabla_x.\nabla_{x'}G_2)$. Using Eq.(7) and the
fact that $\nabla{Q}(\vec{x},t)$ introduces extra power of $q$ in $Q$ and is
thus higher order in $q_c/k_F$, we find that to leading order in $q_c/k_F$, the $Q(\vec{x},t)$ factor does not contribute and $P(\vec{x},t)$ is
given by the usual BCS expression. In particular, the fourier transform $P(\vec{q}\rightarrow0,\omega=0)$
is given by
\[
P(q\rightarrow0,0)=\displaystyle{\frac{1}{L^{2}}\sum_{\vec{k}}\frac{E(k)E(k')-
\epsilon(k)\epsilon(k')-\Delta^{2}}{4E(k)^3}k^{2}|_{\vec{k}'\rightarrow \vec{k}}}, \]
which vanishes when $\Delta\neq0$ indicating that the system exhibits 
Meissner effect as in usual BCS superconductors.

   The apparent contradiction between existence of Meissner effect and 
vanishing of BCS order parameter $\Delta_{BCS}\rightarrow0$ can be understood
in the marginal Fermi liquid picture if we assume that the pairing
amplitude of {\em quasi-particles} $\Delta_{qp}=\Delta$ is nonzero in the 
ground state. The bare fermion pairing amplitude is given by the usual Fermi
liquid relation $\Delta_{BCS}\sim{z}\Delta_{qs}$ which is nonzero for usual Fermi liquids with renormalization factor $z\neq0$. However, $z\sim(q_cL)^{-\gamma}$ and vanishes in the $L\rightarrow\infty$ limit in the present case and $\Delta_{BCS}\rightarrow0$. Nevertheless superconductivity still exists because $\Delta_{qp}\neq0$! Notice that this is a rather general 
consequence of a marginal Fermi liquid superconducting state where pairing occurs between quasi-particles. In particular, the argument is completely independent of the symmetry of the superconducting order parameters. 
Notice that quantum Monte-Carlo studies on $t-J$ model have indicated that
the pairing function of electrons $\Delta_{BCS}$ seems to vanish as the size of
the system increases\cite{tklee}. This result was interpreted as indicating the absence of superconductivity in $t-J$ model\cite{tklee}. Our finding here
suggests an alternative possibility: superconductivity may still exists in
$t-J$ model if the pairing quasi-particles have vanishing wavefunction overlap with bare electrons in the system. 

Summarizing, we have studied in this paper the influence of long range repulsive electron-electron interaction on an s-wave superconductor at two dimensions. We
find that the bare electron BCS pairing function $\Delta_{BCS}$ is strongly suppressed by the plasma oscillation, with however superconductivity (Meissner effect) kept intact. We interpret the system as a marginal Fermi liquid
superconductor where the pairing objects are quasi-particles with vanishing
wavefunction overlap $z^{1\over2}$ with bare electrons in the system. In 
particular, we point out that this situation might have occurred in the $t-J$ 
model where the (d-wave) bare electron pairing function seems to vanish when 
size of system increases\cite{tklee}. We also point out that a pseudo-gap like 
structure in both the normal and superconducting states appears in the tunneling
density of states of bare electrons, as a result of the vanishing wavefunction 
overlap $z\rightarrow0$. An important
property of our microscopic model is that the marginal Fermi 
liquid effect comes from coupling of electrons to (high energy) plasmons and
is not destroyed by superconductivity in contrast with many other 
models\cite{2,3}. The anomalous behaviour we observed in supeconducting state 
is a direct consequence of survival of marginal Fermi liquid effect in the superconducting state and our calculation demonstrates explicitly the existence of this new class of marginal Fermi liquid superconductors.  

   This work is supported by HKUGC through RGC grant HKUST6143/97P.

\end{document}